\documentclass[11pt,twoside]{article}
\usepackage{asp2006}
\usepackage{subfigure}
\usepackage{graphicx}
\usepackage{lscape}
\begin{document}

\title{EXOFIT: Bayesian Estimation of Orbital Parameters of Extrasolar Planets}
\author{Sreekumar T. Balan}
\affil{Astrophysics Group, Cavendish Laboratory, JJ Thomson Avenue, Cambridge CB3 0HE, UK}
\author{Ofer Lahav}
\affil{Department of Physics and Astronomy, University College, Gower Street, London WC1E 6BT, UK}

\begin{abstract}
We introduce \textsc{EXOFIT}, a Bayesian tool for estimating orbital
parameters of extra-solar planets from radial velocity
measurements. \textsc{EXOFIT} can search for either one or two planets
at present. \textsc{EXOFIT} employs Markov Chain Monte Carlo method
implemented in an object oriented manner. As an example we re-analyze
the orbital solution of HD155358 and the results are compared with
that of the published orbital parameters. In order to check the
agreement of the \textsc{EXOFIT} orbital parameters with the
published ones we examined radial velocity data of 30 stars taken
randomly from \textit{www.exoplanet.eu}. We show that while
orbital periods agree in both methods, \textsc{EXOFIT} prefers
lower eccentricity solutions for planets with higher ($e\geq 0.5$)
orbital eccentricities.

\end{abstract}
\keywords{methods: data analysis-methods: numerical-methods: statistical-techniques:radial velocities- planetary systems}

\section{Introduction}
More than a decade of extensive search for extra-solar planets has resulted in nearly 300 planets. Majority of the contribution to the extra-solar planet count comes from radial velocity method. Radial velocity data is traditionally analyzed first by a periodogram \citep{lomb1973, scargle1982} for the orbital period and then for the other orbital parameters using conventional optimization methods. Bayesian methods for the estimation of orbital parameters of extra-solar planets were introduced by \citet{gregory2005a} and \citet{ford2005}. Their work shows that these methods provide us a framework to tackle the problems associated with the traditional methods in a  transparent and robust manner. \textsc{EXOFIT}\footnote{www.star.ucl.ac.uk/$\sim$lahav/exofit.html} of \cite{balan2008} is the first publicly available package for Bayesian estimation orbital parameters from radial velocity measurements. In this article we discuss the application of \textsc{EXOFIT} to the radial velocity data of HD155358 and compare our results with the published orbital solution  \citep{cochran2007}. We also show, by analyzing a randomly selected sample of radial velocity data from \textit{www.exoplanet.eu} that \textsc{EXOFIT} disagrees with the published orbital eccentricities in many instances while the orbital periods from \textsc{EXOFIT} matches closely with the published ones.

The rest of the article is organized as follows. In Section~\ref{modeling} we give a brief introduction to radial velocity modeling. Section~\ref{sec:paraest} describes Bayesian approach to parameter estimation. \textsc{EXOFIT} is introduced in Section~\ref{sec:exofit} and its application to the radial velocity data is discussed in section~\ref{sec:HD155358}. Section~\ref{sec:exofitVSeu} explains the analysis of radial velocity data of a sample of stars taken form \textit{www.exoplanet.eu} using \textsc{EXOFIT}. We conclude this article in Section~\ref{sec:summary} and provide an outline on planned work.

\section{\label{modeling}Modeling of Radial Velocity Data}

Radial velocity data consists of a set of measured radial velocity entries, corresponding time of observation and the uncertainty in each measurement. Observed radial velocity data is modeled by the equation \citep{gregory2005a}
\begin{equation}
\label{dataModel}
d_i=\nu_i+\epsilon_i+\delta,
\end{equation}
where $d_i$ is the measured radial velocity data for the $i$th instant of time $t_i$, $\nu_i$ is true radial velocity of the star, $\epsilon_i$ is the measurement error assigned by the observer and $\delta$ represents any unknown noise present (e.g., signal from another planet) in the data. For a statistician, $\delta$ is a nuisance parameter. The true radial velocity can be simulated using a mathematical model. Disregarding any interactions between planets, the radial velocity of a star for a typical $n$-planet model can be approximately written as a linear combination of $n$ single planet radial velocities. Thus,
\begin{eqnarray}
\label{nPlanetEquation}
 v= V -\sum_{i=1}^{n}K_i\big(\sin(f_{i}+\varpi_i)+e_i\,\sin \varpi_i\big),
\end{eqnarray}
 where $V,K,f,e,w$ represent the systematic velocity, amplitude, true anomaly, eccentricity and the longitude of periastron respectively.  For the full formalism see the User's Guide to \textsc{EXOFIT}. 

\section{Bayesian Parameter Estimation}
\label{sec:paraest}
Bayesian paradigm has its origins in an article published posthumously by Rev. Thomas Bayes in 1763. 
Since then the theorem has played a central part in probabilistic inference. For latest examples in cosmology see \cite{feroz2008} and \cite{lewis2002}. Bayesian methods for the estimation of orbital parameters of extra-solar planets were introduced by \cite{gregory2005a} and \cite{ford2005} and their research show that this approach has an edge over the traditional methods when dealing with for e.g., highly eccentric orbits. These methods also provide a straight forward  way of dealing with nuisance parameters and a robust way of estimating uncertainties associated with the estimates of orbital parameters.

Bayes' theorem for a set of parameters $\mathbf{\Theta}$ in  model $H$ and data $\mathbf{D}$ can be written as, 
\begin{equation}
\Pr(\mathbf{\Theta}|\mathbf{D},H)=\frac{\Pr(\mathbf{D}|\mathbf{\Theta},H)\Pr(\mathbf{\Theta}|H)}{\Pr(\mathbf{D}|H)}.
\end{equation} 
In the above equation $\Pr(\mathbf{\Theta}|\mathbf{D},H)$ is the posterior probability distribution of parameters, 
$\Pr(\mathbf{D}|\mathbf{\Theta},H)$ is the likelihood of the data, $\Pr(\mathbf{\Theta}|H)$ represents the prior probability distribution of the parameters and $\Pr(\mathbf{D}|H)$ is called the Bayesian evidence. For parameter estimation problems we could simply write the above equation as
\begin{equation}
\label{eqn:posteriorProb}
\Pr(\mathbf{\Theta}|\mathbf{D},H) \propto \Pr(\mathbf{D}|\mathbf{\Theta},H)\Pr(\mathbf{\Theta}|H)
\end{equation}

Computing the right hand side of Equation~\ref{eqn:posteriorProb} is the central point of any Bayesian parameter estimation . Analytical solutions can be derived for some special cases. However, in general we use numerical methods to compute the posterior distribution. Although many approximation methods exist, this area is dominated by Markov Chain Monte Carlo (MCMC) and other sampling methods. These rely on the a random walk through the parameter space and make use of the fact that posterior density is proportional to the number of points visited in the volume considered. We calculate the marginal posterior distribution of each parameter by simple plotting a histogram of the final set of samples. 

Bayesian modeling of the problem consists of defining each of the components mentioned in Equation~\ref{eqn:posteriorProb}. This will be discussed from the context of estimation of orbital parameters when we apply \textsc{EXOFIT} to the published radial velocity data of HD155358 in Section~\ref{sec:HD155358}.

\section{\textsc{EXOFIT}}
\label{sec:exofit}
\textsc{EXOFIT} is an easy to use documented software for estimating the orbital parameters from radial velocity measurements and it is freely available. It is based on Bayesian MCMC method and is implemented in an object oriented framework in \textsc{C++}. It can be easily extended to analyze the radial velocity data of more than two planets as well as data from transit photometry. Output of \textsc{EXOFIT} is a set of posterior samples. These can be analyzed using any standard statistical software. This explained in User's Guide which can be download from \textsc{EXOFIT} website\footnote{www.star.ucl.ac.uk/$\sim$lahav/exofit.html}. We also provide sample script for \textsc{R} statistical environment\footnote{www.r-project.org} to analyze posterior samples. Improved parameterization for the problem and novel sampling techniques for \textsc{EXOFIT} are under development. 

\section{Application to Radial Velocity Data of HD155358}
\label{sec:HD155358}
In this section we develop a Bayesian model for the parameter estimation and apply \textsc{EXOFIT} to extract the orbital parameters of the companions of HD155358. We start by defining the likelihood of the data. As mentioned in Section~\ref{modeling} we assume that the data consists of true radial velocity and some noise. The component of noise arising from the known measurement errors $\epsilon_i$ is assumed to be normally distributed with standard deviation $\sigma_i$ for the $i$th entry in the data. The probability distribution for the unaccounted noise component $\delta$ is chosen to a Gaussian distribution with finite variance $s^2$. Therefore, the distribution of the combination $\epsilon_i+\delta$ can be considered as a Gaussian with a variance of $\sigma_i^2+s^2$.

Assuming each measurement error $\epsilon_i$ to be independent and since they follow a Gaussian distribution, the likelihood of data can be written as a product $N$ Gaussians\citep{gregory2005a} where $N$ is the number of entries in the data. Therefore,
\begin{equation}
\label{eqn:lik}
\Pr(\mathbf{D}|\mathbf{\Theta},H)=A\, \exp \Biggl[-\sum_{i=1}^{N}\frac{(d_i-v_i)^2}{2(\sigma_i^2+s^2)}\Biggr]\,,
\end{equation} 
where
\begin{equation}
A=(2\pi)^{-N/2}\Biggl[\prod_{i=1}^{N}\big(\sigma_i^2+s^2\big)^{-1/2}\Biggr]\,.
\end{equation}
Thus, our parameter space is $\{V,T_1,K_1,e_1,w_1,\chi_1,T_2,K_2,e_2,w_2,\chi_2,s\}$, first 11 from the mathematical model and last one representing the nuisance parameter $\delta$. The parameter is $\chi$ is defined for computational purposes and marks the periastron passage time as function period $T$. For more details please consult \textsc{EXOFIT} User's Guide.  

Table~\ref{tab:prior12}, taken from \cite{balan2008} gives the prior probability distributions of each parameter in the model. These priors are chosen in such a way that they allow likelihood term in the Equation~\ref{eqn:posteriorProb} to dominate the posterior distribution and thus ensuring inference to be drawn from the observed data.
\begin{table}[!ht]
\scriptsize
\caption[Choice of Priors for the 2-planet model]
{The assumed  prior distribution of orbital parameters and their boundaries for a 2-planet model.}
\smallskip
\begin{center}
\begin{tabular}{cccccc}
\tableline
\noalign{\smallskip}
 Para. & Prior  & Mathematical Form & Min & Max \\ 
\hline
$V(ms^{-1})$ & Uniform  & $\frac{1}{V_{max}-V_{min}}$ & -2000 & 2000 \\ 
 $T_1(days)$& Jeffreys & $\frac{1}{T_1\,\ln\Big(\frac{T_{1\,max}}{T_{1\,min}}\Big)}$ & 0.2 & 15000 \\ 
$K_{1}(ms^{-1})$ & Mod. Jeffreys & $\frac{(K_1+K_{1\,0})^{-1}}{\ln\big(\frac{K_{1\,0}+K_{1\,max}}{K_{1\,0}}\big)}$ & 0.0 &2000 \\ 
 $e_1$ & Uniform & 1 & 0 & 1 \\ 
 $\varpi_1$ & Uniform & $\frac{1}{2\pi}$ & 0 & $2\pi$ \\ 
 $\chi_1$ & Uniform & 1 & 0 & 1 \\ 
 $T_2(days)$& Jeffreys & $\frac{1}{T_2\,\ln\Big(\frac{T_{2\,max}}{T_{2\,min}}\Big)}$ & 0.2 & 15000 \\ 
 $K_{2}(ms^{-1})$ & Mod. Jeffreys &  $\frac{(K_2+K_{2\,0})^{-1}}{\ln\big(\frac{K_{2\,0}+K_{2\,max}}{K_{2\,0}}\big)}$ & 0.0 &2000 \\ 
 $e_2$ & Uniform & 1 & 0 & 1 \\ 
$\varpi_2$ & Uniform & $\frac{1}{2\pi}$ & 0 & $2\pi$ \\ 
$\chi_2$ & Uniform & 1 & 0 & 1 \\ 
$s(ms^{-1})$ & Mod. Jeffreys & $\frac{(s+s_{0})^{-1}}{\ln\big(\frac{s_{0}+s_{max}}{s_{0}}\big)}$ & 0 & $2000$\\
\noalign{\smallskip}
\tableline
\end{tabular}
\end{center}
\label{tab:prior12}
\end{table}

\begin{figure}[!ht]
\subfigure[Marginal posterior destributions]{
\label{fig:HD155358dens}
\includegraphics[scale=0.3]{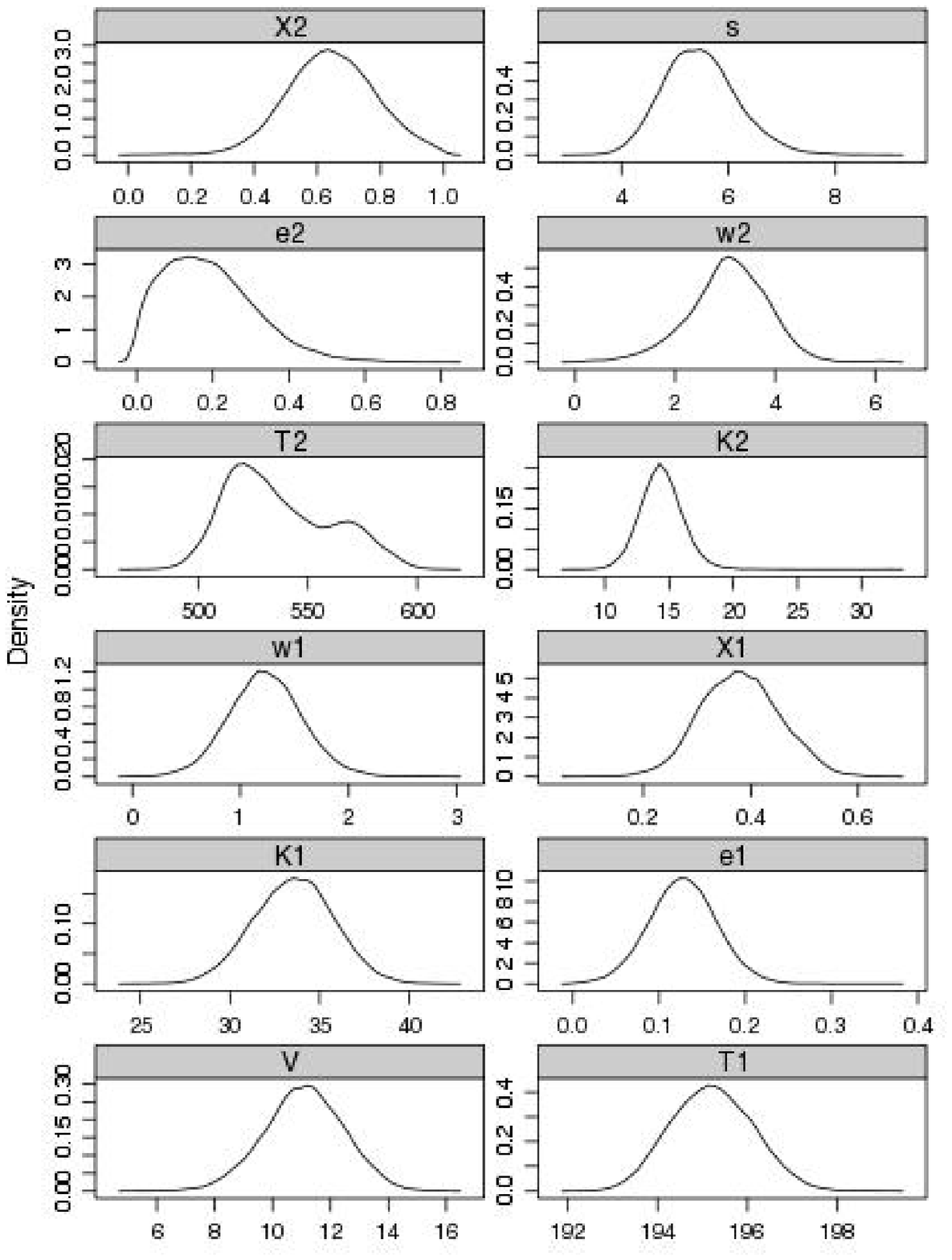}
}
\subfigure[Radial velocity curve]{
\label{fig:HD155358orb}
\includegraphics[scale=0.3]{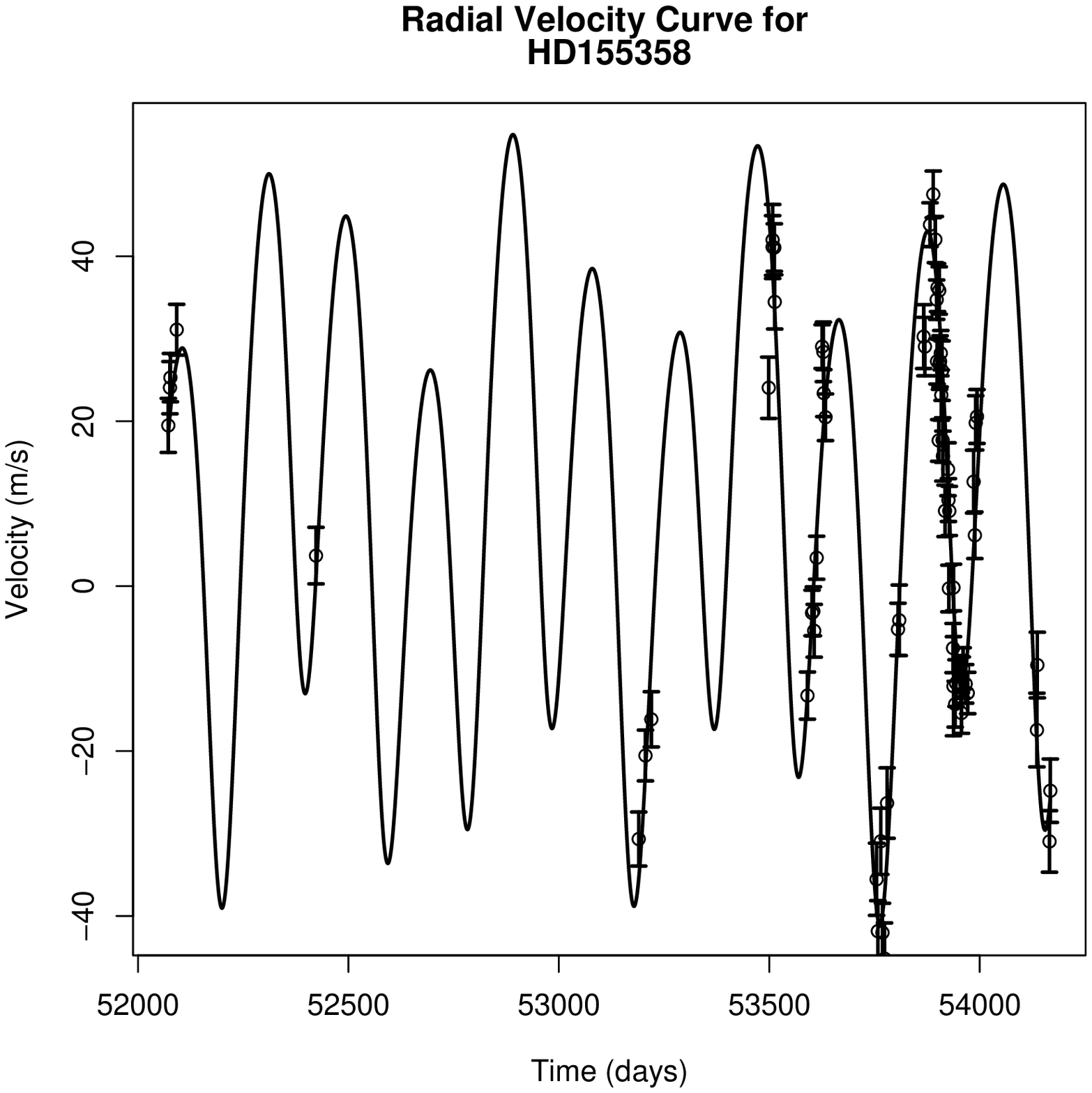}
}
\caption{Marginal postrior distribuotions of orbital orbital parameters of the companions to HD155358 are shown in the left panel. Right panel shows the corresponding radial velocity curve for HD155358.}
\label{fig:HD155358Solutions}
\end{figure}

Application of \textsc{EXOFIT} revealed the posterior distribution of planets as given in Figure \ref{fig:HD155358dens} and the corresponding radial velocity curve is shown in Figure~\ref{fig:HD155358orb}. We compare our results to the published results by \citep{cochran2007} in Table \ref{tab:exoVSpub}. Although our results looks similar, we notice a noise factor($s$) of $5.43\,ms^{-1}$ which indicates the presence of an additional signal in the data. We did look for a third planet in the system, but our results were inconclusive. This issue can be settled if we have more observations down the line.

\begin{table}[!ht]
\caption[Comparison of orbital parameters]
{A table showing the summary of the posterior distribution of orbital parameters extracted with \textsc{EXOFIT} 2-planet model and the published orbital solution by \cite{cochran2007}. Columns 2, 3 and 4 show
the posterior mean (and standard deviation), median (and 25\% and 75\%
quantiles) and the maximum a posteriori, i.e. posterior mode (and 68.3\% highest density
regions) respectively. Mass of the star was assumed to be $0.87M_{\oplus}$.}
\smallskip
\begin{center}
\scriptsize
\begin{tabular}{ccccc}
\tableline
\noalign{\smallskip}
Parameters	&	$\textsc{EXOFIT}$(Mean)			&	$\textsc{EXOFIT}$(Median)					&	$\textsc{EXOFIT}$(Mode)					&	Cochran et al.			\\
\tableline
\noalign{\smallskip}
$V(ms^{-1})$	&$	11.09	\pm	1.37	$&$	11.11	_{-	0.91	}^{+	0.91	}$&$	11.23	_{-	1.32	}^{+	1.17	}$&$				$\\
\noalign{\smallskip}
$T_1(days)$	&$	195.22	\pm	0.91	$&$	195.20	_{-	0.63	}^{+	0.64	}$&$	195.17	_{-	0.84	}^{+	0.87	}$&$	195	\pm	1.1	$\\
\noalign{\smallskip}
$K_1(ms^{-1})$	&$	33.56	\pm	2.22	$&$	33.59	_{-	1.55	}^{+	1.48	}$&$	33.58	_{-	2.02	}^{+	2.07	}$&$	34.6	\pm	3	$\\
\noalign{\smallskip}
$e_1$	&$	0.13	\pm	0.04	$&$	0.13	_{-	0.03	}^{+	0.03	}$&$	0.13	_{-	0.04	}^{+	0.03	}$&$	0.112	\pm	0.037	$\\
\noalign{\smallskip}
$\varpi_1(degrees)$	&$	160.78	\pm	19.73	$&$	160.46	_{-	12.77	}^{+	12.82	}$&$	158.20	_{-	15.33	}^{+	19.30	}$&$	162	\pm	20	$\\
\noalign{\smallskip}
$\chi_1$	&$	0.38	\pm	0.07	$&$	0.38	_{-	0.05	}^{+	0.05	}$&$	0.38	_{-	0.07	}^{+	0.06	}$&$				$\\
\noalign{\smallskip}
$T_2(days)$	&$	537.05	\pm	24.00	$&$	532.15	_{-	13.84	}^{+	22.78	}$&$	519.77	_{-	13.46	}^{+	29.19	}$&$	530.3	\pm	27.2	$\\
\noalign{\smallskip}
$K_2(ms^{-1})$	&$	14.40	\pm	1.68	$&$	14.34	_{-	1.03	}^{+	1.07	}$&$	14.21	_{-	1.35	}^{+	1.52	}$&$	14.1	\pm	1.6	$\\
\noalign{\smallskip}
$e_2$	&$	0.19	\pm	0.12	$&$	0.18	_{-	0.08	}^{+	0.09	}$&$	0.14	_{-	0.10	}^{+	0.11	}$&$	0.176	\pm	0.174	$\\
\noalign{\smallskip}
$\varpi_2(degrees)$	&$	265.62	\pm	45.42	$&$	267.53	_{-	28.62	}^{+	28.16	}$&$	265.28	_{-	32.34	}^{+	45.48	}$&$	279	\pm	38	$\\
\noalign{\smallskip}
$\chi_2$	&$	0.64	\pm	0.14	$&$	0.64	_{-	0.09	}^{+	0.10	}$&$	0.63	_{-	0.11	}^{+	0.14	}$&$				$\\
\noalign{\smallskip}
$s(ms^{-1})$	&$	5.47	\pm	0.69	$&$	5.44	_{-	0.45	}^{+	0.47	}$&$	5.48	_{-	0.70	}^{+	0.52	}$&$				$\\
\noalign{\smallskip}
$Tp_1(BJD)$	&$	2453946.95	\pm	13.93	$&$	2453946.95	_{-	10.00	}^{+	10.00	}$&$	2453950.63	_{-	13.90	}^{+	14.26	}$&$	2453950	\pm	10.4	$\\
\noalign{\smallskip}
$M_1\sin\,i$	&$	0.87	\pm	0.06	$&$	0.87	_{-	0.04	}^{+	0.04	}$&$	0.87	_{-	0.06	}^{+	0.06	}$&$	0.89	\pm	0.12	$\\
\noalign{\smallskip}
$a_1(AU)$	&$	0.63	\pm	0.00	$&$	0.63	_{-	0.00	}^{+	0.00	}$&$	0.63	_{-	0.00	}^{+	0.00	}$&$	0.628	\pm	0.02	$\\
\noalign{\smallskip}
$Tp_2(BJD)$	&$	2454408.22	\pm	70.22	$&$	2454408.22	_{-	50.00	}^{+	50.00	}$&$	2454420.27	_{-	76.26	}^{+	60.57	}$&$	2454420.3	\pm	79.3	$\\
\noalign{\smallskip}
$M_2\sin\,i$	&$	0.51	\pm	0.06	$&$	0.51	_{-	0.04	}^{+	0.04	}$&$	0.51	_{-	0.06	}^{+	0.06	}$&$	0.504	\pm	0.075	$\\
\noalign{\smallskip}
$a_2(AU)$	&$	1.23	\pm	0.04	$&$	1.23	_{-	0.02	}^{+	0.03	}$&$	1.21	_{-	0.02	}^{+	0.07	}$&$	1.224	\pm	0.081	$\\
\noalign{\smallskip}
\tableline
\end{tabular}
\end{center}
\label{tab:exoVSpub}
\end{table}
\normalsize
\section{\textsc{EXOFIT} vs exoplanet.eu}
\label{sec:exofitVSeu}
\begin{figure}[!ht]

\subfigure[Log(Period)]{
\includegraphics[scale=0.25]{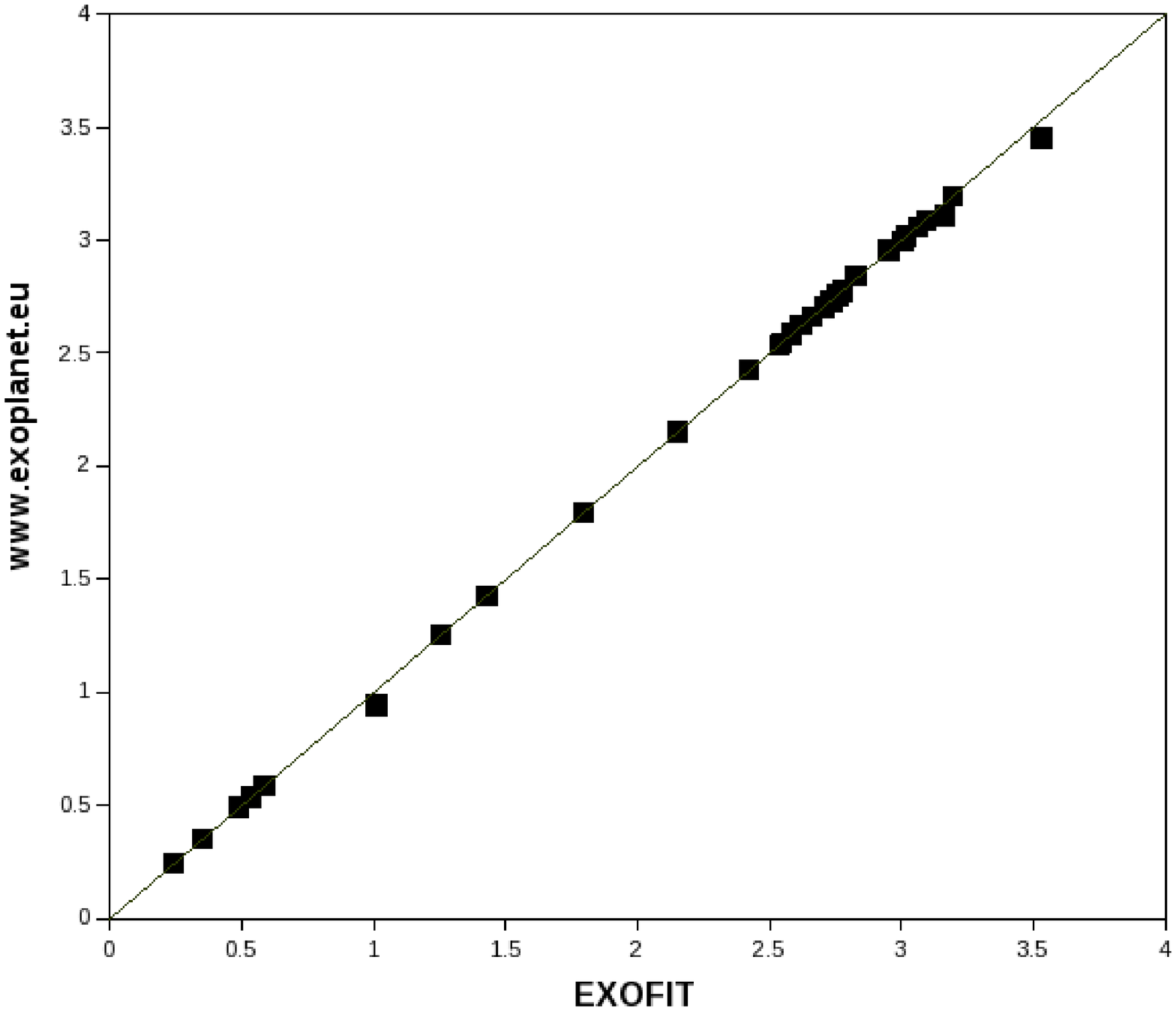}
}
\hspace{.1in}
\subfigure[Eccentricities]{
\includegraphics[scale=0.25]{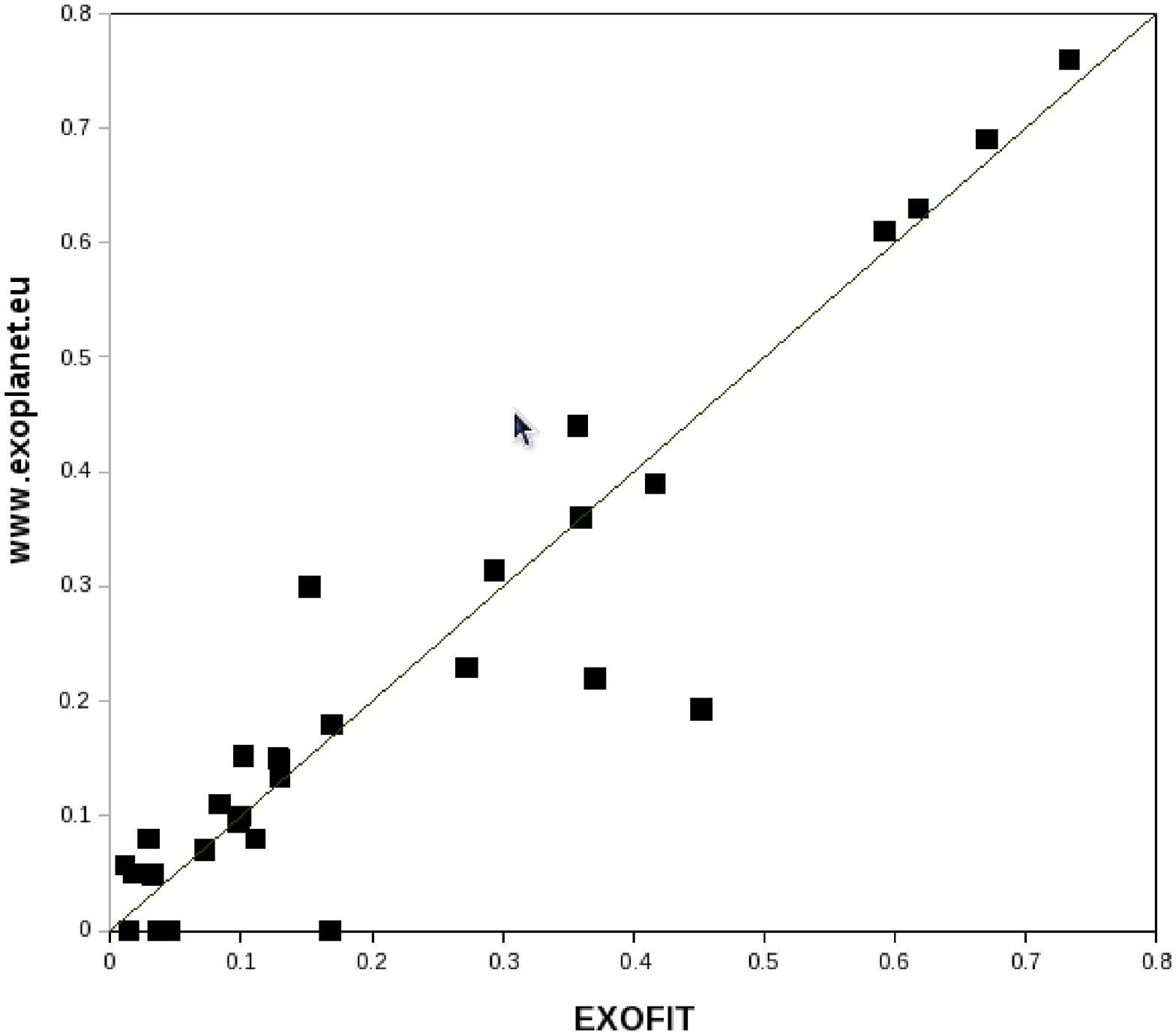}
}
\caption{Figure on the left plots $\log$(orbital period) of extra-solar planets, with results obtained by \textsc{EXOFIT} on the horizontal axis and published ones on the vertical axis. Figure on the right shows a comparison of orbital eccentricities with results obtained from \textsc{EXOFIT} on the horizontal axis and published ones on the vertical axis. }
\label{fig:exoVSeu}
\end{figure}

In this section we compare the published orbital periods and eccentricities of 30 extra-solar planets taken randomly from \textit{www.exoplanet.eu} to that obtained by analyzing corresponding radial velocity data  with \textsc{EXOFIT} .  The  results are summarized in Figure~\ref{fig:exoVSeu}. It can be observed that the orbital periods extracted with \textsc{EXOFIT} matches closely with published ones. However, the eccentricities show apparent variation from the published values.  In fact \textsc{EXOFIT} tends to obtain lower eccentricity solutions. This is clearly noticeable for planets with eccentricities greater that 0.5 (i.e. $\simeq$ 10\% of the planet population). Median was used as the point estimator for marginal posterior distribution orbital parameters of exo-planets extracted with \textsc{EXOFIT} while comparing with the published results. This suggests that the orbital eccentricity is poorly constrained in many occasions because of the sparse sampling of the data points.

\section{Summary and Future Work}
\label{sec:summary}
We have introduced \textsc{EXOFIT} by reanalyzing the orbital solution of HD155358 by \cite{cochran2007}. \textsc{EXOFIT} provides a full Bayesian analysis of the problem and spits out the marginal posterior distribution of orbital parameters. We have also compared the orbital parameters obtained by \textsc{EXOFIT} to the published orbital solutions of 30 extra-solar planets taken randomly from \textit{www.exoplanet.eu}, the objective being the reanalysis of the data using a single method  and the estimation of corresponding orbital parameters and their uncertainties. The results show that while the orbital periods from \textsc{EXOFIT} agrees closely with the published ones, the eccentricities show significant variation from the published results. This fact indicates that orbital eccentricity is not accurately constrained and this degeneracy should be considered more carefully. 

We plan to improve to the efficiency of \textsc{EXOFIT} by considering new parameterization for the problem as well as faster sampling techniques. Bayesian model selection will also be considered for the future versions of \textsc{EXOFIT}. We intend to extend our analysis to more planets and planetary data from transit photometry to provide a comprehensive Bayesian analysis of the statistical properties of orbital parameters of extra-solar planets.   

\acknowledgements SB and OL would like to thank the organizers of the \textit{Molecules 2008} for giving the opportunity to attend the conference. SB would like the thank the organizing committee for the financial support provided. 
OL acknowledges the support of a Royal Society Wolfson Research Merit Award.


\end{document}